\documentclass[aps,prb,superscriptaddress,showpacs,twocolumn]{revtex4}%
\usepackage{amsfonts}
\usepackage{amsmath}
\usepackage{amssymb}
\usepackage{graphicx}%
\providecommand{\U}[1]{\protect\rule{.1in}{.1in}}
\begin{document}
\title{Quantized spin pump on helical edge states of a topological insulator}

\author{Mei-Juan Wang}
\affiliation{Department of Physics, Southeast University, Nanjing, 210096, China}

\author{J. Wang}
\affiliation{Department of Physics, Southeast University, Nanjing, 210096, China}
\author{Jun-Feng  Liu}
\affiliation{Department of Physics, South University of Science and Technology of China, Shenzhen 518055,  China}
\email{liujf@sustc.edu.cn}

\begin{abstract}
We report a theoretical study of the quantized spin pump in a traditional two-parameter quantum pump device that is based on the helical edge states of a quantum spin Hall insulator. By introducing two time-dependent magnetizations out of phase as the pumping parameters, we found that when the Fermi energy resides in the energy gap opened by magnetization, an integer number of charges or spins can be pumped out in a pumping cycle and ascribed to the possible topological interface state born in between the two pumping potentials. The quantized pump current can be fully spin-polarized, spin-unpolarized, or  pure spin current while its direction can be abruptly reversed by some system parameters such as the pumping phase and local gate voltage. Our findings may shed light on generation of a quantized spin pump.
\end{abstract}


\maketitle
\section{Introduction}

Quantum parametric pump like the Archimedean screw that can pump water by a rotating spiral tube is a striking topic in the context of quantum transport through nanostructures and arises in nature from the geometric Berry phases and quantum interference effects\cite{alt,1,2,3}. Usually, the quantum parametric pump is implemented through two or more time-dependent potentials or perturbations out of phase in mesoscopic devices and can produce a DC current without any external bias, which is proportional to the geometric area encircled by time-dependent parameters\cite{3} in the adiabatic limit.

One of ultimate goals in the field of quantum parametric pump is to find a quantized charge pump that in a cyclic period, an integer number of charges are pumped out to flow through the device.  It is argued that the quantized charge pump can revolutionize electrical metrology by enabling the ampere to be redefined in terms of the elementary charge of an electron\cite{4,5,6,7}. In condensed matter experiments, such a quantized pump were demonstrated in the quantum dot system with the help of strong Coulomb interaction of electrons\cite{8,9,10,11,12,13,14}. In the noninteracting system, a celebrated proposal of quantized pump is the Thouless topological pump\cite{15} in which a one-dimensional (1D) moving potential can pump out integral charges in a pump cycle, when the Fermi energy lies in the energy gap opened by the moving potential. Actually, each pump cycle transports integral electronic charges, and the integer is uniquely determined by a topological invariant: the Chern number of the quantum system\cite{16} which is defined through dimension extension in the 1D system. Certainly, topological charge pumping can be understood as a dynamical analog of the integer quantum Hall effect\cite{17,18}: the pumped charge can be mapped exactly to the quantized Hall conductance of a two-dimensional electronic system.

Very recently, several groups\cite{19,20,21} have independently measured the topological pump in 1D optical superlattice systems due to the advances in constructing optical lattice structures. However, it is still a big challenge for realizing such a topological pump in condensed-matter experiments, because the creation of a dynamical superlattice potential critically relies on the presence and control of superimposed oscillating local voltages\cite{22}. Therefore, a simpler and more practical theory is currently desirable for performing such a quantized pump in noninteracting electron systems not merely limited to this 1D Thouless topological pump as well as its variations\cite{23,24,25,26,27,28,29}.

In a previous work\cite{30}, authors proposed a quantized pump model based on the traditional two-parameter pump protocol in the graphene system and showed that two time-dependent staggered potentials with a phase lag as pumping parameters can result in a quantized charge pump effect. The key point is that the pumping potentials introduced can open an energy gap of massless Dirac electrons of graphene. Since the staggered potentials are very difficult to be operated in graphene, and the massless Dirac electrons are ubiquitous in the edge or surface states of a topological insulator, we in this work investigate the possible quantized parametric pump effect by utilizing such topological edge states, which can, in principle, be gapped by introducing some interaction breaking the symmetry that protects the original topological state. A typical example is the 1D helical edge state of a two-dimensional (2D) quantum spin Hall insulator (QSHI)\cite{31,32,33}: when the magnetization breaking time reversal symmetry is considered, the edge states would be gapped\cite{31,33,34} as long as the magnetization direction is not parallel to the intrinsic spin direction of helical edge states. We will show that two time-dependent magnetic materials\cite{35,36,37,38,39} with a phase lag in between them,  like the AC magnetic field or precessing ferromagnets, can give rise to a quantized charge or spin pump\cite{40,41,42}, which depends on different magnetization configurations. The quantized charge or spin current can be modulated by the system parameters such as the pumping phase and the local gate voltage. An abrupt current reversal effect of the pumped current, which is quite useful in fabricating quantum switch devices\cite{43}, is also demonstrated.

This work is organized as follows. In Sec.~II, we present a lattice model to calculate the pump current in two pump models: one is the magnetization covering the whole QSHI material, and the other is the magnetization covering only one boundary of QSHI. In Sec.~III, a continuum model is also employed to analyze the obtained numerical results. A model for pure spin pump is further studied in Sec.~IV and a conclusion is drawn in the last section.

\begin{figure}[ptb]
\centering
\includegraphics[bb=0 0 255 193, width=3.0in]{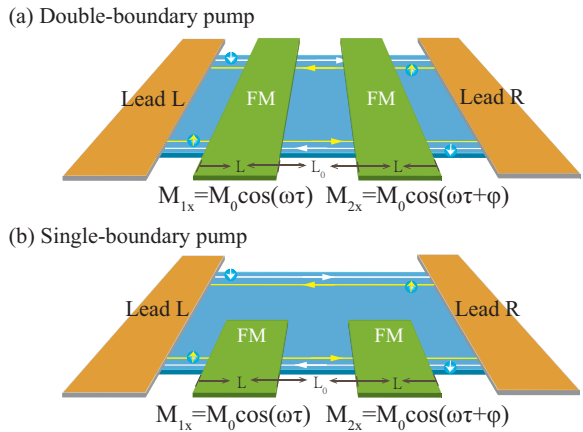}
\caption{(Color online) Schematic of two pump devices based on the helical states of QSHI.  One is the pump potentials of two FM islands covering double boundaries of QSHI (a) and the second one is the FM deposited only one boundary of QSHI (b). The spin-momentum locked electrons circulate along the boundary of QSHI and two FM magnetizations, $M_{1x}$ and $M_{2x}$, evolve with time adiabatically. The pumped current is assumed to flow through the two contacted leads. $L$ and $L_0$ stand for the length of two FM islands and distance between them, respectively.}
\end{figure}

\section{Lattice model}
We consider typical two-parameter pump devices based on the 2D QSHI as schematically shown in Fig.~1, where two ferromagnetic (FM) islands are deposited either on the whole QSHI covering two helical edge states [Fig.~1(a)] or on one boundary of QSHI [Fig.~1(b)]. The former is referred to as the double-boundary pump device, and the later is dubbed as the single-boundary one. The setup is assumed to contact outside world through the left and right leads without any applied bias. It is assumed that the two FM magnetizations are taken as the pumping parameters varying with time adiabatically, and there is an onset phase difference $\varphi$ between them. For the 2D QSHI, both the Kane-Mele\cite{31} and the Bernevig-Hughes-Zhang\cite{32} models are suitable for the study purpose of this work, and the final pump results are almost the same. Thus, the former case is adopted here and the device Hamiltonian in a lattice version is given by
\begin {eqnarray}
{\cal{H}}=&-t\sum_{\langle{ij}\rangle\sigma}C_{i\sigma}^{\dagger}C_{j\sigma} +\frac{\lambda_{so}}{3\sqrt{3}}\sum_{\ll{ij}\gg\sigma}\upsilon_{ij}C_{i\sigma}^{\dagger}s_{z}C_{j\sigma} \\ \nonumber
&+\sum_{i\beta\gamma}C_{i\beta}^{\dagger}({\boldsymbol{\sigma}}\cdot\mathbf{M}_\tau)C_{i\gamma}.
\end{eqnarray}
Here, the first term describes  pristine graphene, $\langle{ij}\rangle$ stands for the nearest-neighboring sites, $C_{i\sigma(\beta,\gamma)}^{\dagger}(C_{i\sigma(\beta,\gamma)})$ is the creation (annihilation) operator at site $i$ with spin $\sigma(\beta,\gamma)$, and $t$ is the hopping energy of electrons. The second term is the spin-orbit interaction accounting for the topological phase in graphene with its strength, $\lambda_{so}$,  $s_z$ is the spin operator, $\ll{ij}\gg$ represents the next-nearest neighboring sites, and $\upsilon_{ij}=1$ if the next-nearest neighboring hopping is counterclockwise, and $\upsilon_{ij}=-1$ if it is clockwise with respect to the normal of the 2D sheet; the third term denotes the spin exchange energy with $\mathbf{M}_\tau$ ($\tau$, time argument) being the time-depended magnetization on each site $i$, which is assumed uniform in the FM island regions but vanishing outside of FMs.

The quantum spin axis is set along the intrinsic spin eigendirection of spin orbit interaction (or the $z$ direction here), and the direction of $\mathbf{M}_\tau(M_x, M_y)$  is limited in the $xy$ plane, so that it can gap the helical edge states of QSHI. Without loss of generalization, the magnetization is assumed along the $x$ direction, and the pumping phase difference $\varphi$ is considered in the right pumping potential: $M_{1x}=M_0\cos\omega\tau$ and $M_{2x}=M_0\cos(\omega\tau+\varphi)$, where the pumping frequency $\omega$ is infinitesimal, so that the evolving system is justified to keep in the ground state, and $M_0$ is the pumping strength. It is noted here that $M_0$ is considered to be less than the strength of spin orbit interaction, $M_0<\lambda_{so}$, because $\lambda_{so}$ represents the bulk energy gap of QSHI and in our model of quantized spin pump, only the electrons in the helical edge states are assumed active in the pump process. The bulk states of QSHI should be excluded for they are not expected to cause any quantized pumping effect.

Since we focus on the adiabatic pump, the B\"{u}ttiker-Pr\^{e}re-Thomas formula\cite{44} is employed to calculate the pump curent
\begin{equation}
I_{\alpha\sigma}=\frac{ie}{2\pi{T}}\int_{0}^{T}{d\tau}{\text{Tr}}\left({\frac{\partial \mathcal{S}_{\tau}}{\partial\tau }\mathcal{S}_{\tau}^{*}} \right)_{\alpha\sigma,\alpha\sigma},
\end{equation}
where $\mathcal{S}_{\tau}$ is the instantaneous scattering matrix with $\alpha$ being the left or right lead index, $\alpha=L,R$, and $T = 2\pi/\omega$ is the pump cycle. In order to conveniently carry out numerical calculations in a lattice model, the above equation can be modified as\cite{30}
\begin{equation}
I_{\alpha\sigma}=\frac{e}{2\pi{T}}\oint_{0}^{T}{d\tau}
{\text{Tr}}\left(\Gamma G_\tau^r\dot{\mathbf{M}}_{\tau}G_\tau^a\right)_{\alpha\sigma,\alpha\sigma},
\end{equation}
where $\Gamma_{\alpha\sigma}$ is the line-width matrix of the Lead $\alpha$ with spin $\sigma=\uparrow,\downarrow$ and is determined by time-independent Hamilton of QSHI. $G_{\tau}^{r(a)} =[E\pm i0^{+}-{\cal{H}}(\tau)]^{-1}$ is the instantaneous retarded (advanced) Green's function of the two-terminal device, $\dot{\mathbf{M}}_{\tau}= d\mathbf{M}_\tau/d\tau$ is the time derivative of  pump potentials, and the trace is over the transverse sites of a unit slice of the lattice pump model. The Green's function $G^{r(a)}_\tau$ can be calculated by using usual recursive Green's function method since the model device can be decomposed into three parts of left and right leads as well as the scattering region.

\begin{figure}[ptb]
\centering
\includegraphics[bb=0 0 446 376, width=3.0in]{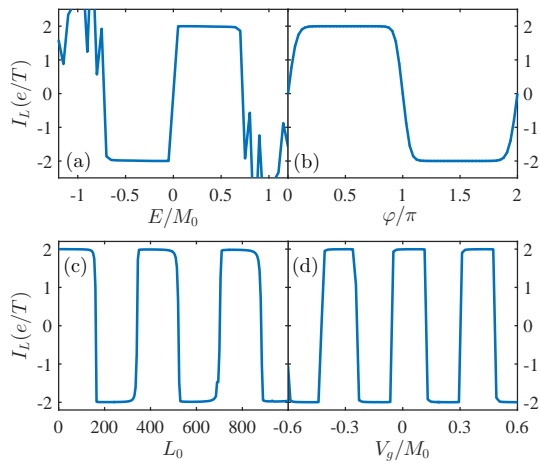}
\caption{(Color online) Pumped charge current $I_L$ as functions of (a) the Fermi energy $E$, (b) the pumping phase $\varphi$, (c) the distance $L_0$, and (d) the local potential $V_g$. Parameters are $\varphi=\pi/2$, $E=0.001t$, $L_0$=0, $V_g=0$, and $M_0=0.01t$.}
\end{figure}

In numerics, a rectangle graphene lattice of QSHI is taken into account here, and the width of device is denoted by the number of zigzag chains of lattice, $N=64$. The sizes of two FM islands are set as the same, $La$, and the distance between them is measured by $L_0a$, where $a$ is the lattice constant of graphene. In calculations, we take the hopping energy $t=1$ eV as the energy unit, the pumping strength is $M_0=0.01t$, the spin orbit interaction strength is  $\lambda_{so}=0.1t$. When the two FM islands mantle only one boundary of QSHI in Fig.~1(b), they are assumed to merely extend into the middle of the QSHI lattice, a half width of the rectangle QSHI.

We first focus on the double-boundary pump device in Fig.~1(a), and the pumped charge current flowing through the left Lead, $I_L=I_{L\uparrow}+I_{L\downarrow}$, is computed according to Eq.~(3).  The pumped current versus the Fermi energy is plotted in Fig.~2(a), and it is clearly shown that $I_L$ fulfills the particle-hole antisymmetry $I_L(E)=-I_L(-E)$, which is a typical property of the two-parameter charge pump device. This reflects the underlying physics that the quantum parametric pump is originated from the interference of different particle-hole particles excited by the pumping potentials\cite{45}. Around $E=0$, $I_L$ is quantized: $I_L=\pm{2e/T}$, where '$2$' stems from the spin degeneracy, i.e, each helical edge state of QSHI at two opposite boundaries should contribute to a charge pumping with opposite spins. Actually, both two opposite helical edge states involved in the pumping process together can simply make the original chirality of electrons disappear.

The quantized pump current in Fig.~2 agrees with the previous conclusion\cite{30} that the pumping results in a two-parameter pump device would be quantized if pumping potentials could open an energy gap of the massless Dirac electrons. Here, our studied model obviously meet these two requirements: $M_{1x}$ and $M_{2x}$  can gap helical edge states, and the original energy dispersion of electrons is of massless Dirac-electron type. It is pointed out that in our pump scheme the local energy gap in  the $M_{1x}(\tau)$ or $M_{2x}(\tau)$ region may close at some special instantaneous time, and only a phase lag $\varphi$ between them would keep the pump device insulating in the whole pumping cycle, $\omega\tau\in(0,2\pi$). So there is an effective global energy gap, $E_{ef}=M_0\sqrt{(1-\cos\varphi)/{2}}$, because  $M_{1x}$ or $M_{2x}$ opens and closes the energy gap asynchronously when they vary with time. The pumped current could be quantized only if the Fermi energy resides in this energy gap $E<E_{ef}$. Thus, it is not strange that the quantized value ($I_L=\pm{2e/T}$) should begin with $E\sim{M_0/\sqrt{2}}$ but not with $E\sim{M_0}$ as numerically shown in Fig.~2(a) when $\varphi=\pi/2$. Similarly, the pump quantization  is attributed to the time-dependent evolution of the possible topological surface state that bridges the two FM islands. As is known, the spin exchange energy in the Hamiltonian of Eq.~(1) can be regarded as a mass term of the Dirac electrons of edge states, so a topological interface state would be born in real space between these two FMs when the signs of $M_{1x}$ and $M_{2x}$  are different at some instantaneous time $\tau$. Oppositely, the same signs of them do not give rise to any interface state. In a complete pumping cycle, its appearance or disappearance brings about an integral number of electrons flowing out of the system.

In terms of the Brouwer's theory\cite{3}, the two-parameter pumping current in the adiabatic limit fulfils the current-phase relationship, $I\sim \sin\varphi$. In Fig.~2(b), $I_L$ versus $\varphi$ is plotted. It is clearly shown that $I_L$ severely deviates from the sine behavior, and instead it exhibits an abrupt current reversal effect from positive quantized value to minus one. $I_L$ is not quantized only when $\varphi\sim{n\pi}$ ($n$ is an integer), because for this situation the effective energy gap approaches to vanishing, $E_{ef}\sim0$,  and the quantization  prerequisite $E<E_{ef}$ can be hardly satisfied. As mentioned above, $I_L$ is determined by the quantum interference effect, so that the dynamic phase of electrons can be employed to control the pumping results. In Fig.~2(c), $I_L$ is depicted as a function of $L_0$, and similarly it displays an abrupt current reversal effect between the two quantized values, $+2e/T$ and $-2e/T$. Actually, one can also use a local gate voltage replacing variation of $L_0$ to modulate $I_L$ as shown in Fig.~2(d), since the gate voltage $V_g$ will change the local wavevector of electrons in helical edge states so as to alter their dynamic phases. Certainly, the later situation is convenient for experimental observations and moreover this abrupt current reversal effect shall have some application potential in quantum switch devices\cite{43}. In calculations, the uniform static potential $eV_g$ is only considered in the nonmagnetic region ($L_0$) between the two FMs in Fig.~1.

\begin{figure}[ptb]
\centering
\includegraphics[bb=0 0 446 376, width=3.0in]{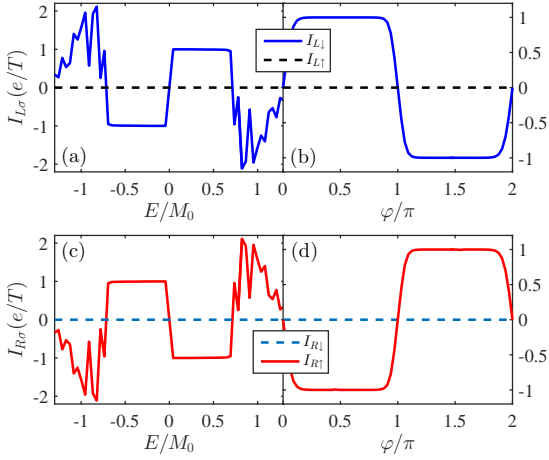}
\caption{(Color online) Spin-dependent pumped current $I_{L\sigma}$ versus (a) the Fermi energy $E$ and (b) the pumping phase $\varphi$. The countpart of the pumped current flowing through right lead $I_{R\sigma}$ are shown in (c) and (d). Parameters are $\varphi=\pi/2$, $E=0.001t$, $L_0$=0, $V_g=0$, and $M_0=0.01t$.}
\end{figure}

We turn to study the single-boundary pump device in Fig.~1(b), where only one edge of QSHI is covered by FMs. $I_L$ as functions of $E$ and $\varphi$ are shown in Fig.~3(a) and 3(b), respectively. It is seen that only one spin-species (say, down spin) current is nonzero and the quantized value is now halved, $I_{L\downarrow}=\pm{e/T}$. But the opposite spin current is prohibited, $I_{L\uparrow}=0$, so the pumped current is fully spin-polarized. This is due to the definite chirality of electrons in helical edge states. In other words, the two-parameter pump device here can extract spin from QSHI. The spin-resolved pumped current flowing into the right lead, $I_R$, is also presented in Fig.~3(c) and 3(d), from which one can find that both current and spin directions are reversed, i.e., the charge current is conserved, $I_L+I_R=0$, but the spin current is nonconserved $I_{Ls}+I_{Rs}\neq0$ ($I_{\alpha{s}}=I_{\alpha\uparrow}-I_{\alpha\downarrow}$). This situation is similar to the uniform magnetization precession on the single boundary of QSHI\cite{38,39}: one spin flowing into the pumping region from left or right lead experiences a flip and then flows into the opposite lead due to the limitation of the electron chirality of helical edge states.

\section{Continuum Model}
It is seen that the two-boundary pump is just a mathematic summation of two single-boundary pumps with oppositely edges of QSHI, i.e., the upper and lower boundaries of QSHI (in Fig.~1) are independently contributing to the pumped current. In this section, we  employ a simple continuum model to further confirm the above numerical calculations.  The pump device based on the 1D helical edge state can be described by the following  Hamiltonian
\begin{eqnarray}
{\cal{H}}=\hbar{v_F}(\eta_z{\sigma_zk_x})+M_{1x}\sigma_x\Theta_1(x)
+M_{2x}\sigma_x\Theta_2(x),
\end{eqnarray}
where the first term is the massless Dirac equation describing the helical edge state, $\eta_z=\pm1$ stands for the opposite chirality of helical edge states, $\sigma_{x,y,z}$ is the real spin Pauli operator, and $k_x$ is the 1D momentum. The second and third terms are the two time-dependent magnetizations, whose direction are fixed along the $x$ axis. $\Theta_1(x)=\Theta(x)\Theta(L-x)$, and $\Theta_2(x)=\Theta(x-L_0-L)\Theta(2L+L_0-x)$ with $\Theta(x)$ being a Heaviside step function.

We directly utilize the B\"{u}ttiker-Pr\^{e}re-Thomas formula\cite{44} of Eq.~(2) for pumped currents, in which the scattering coefficients can be obtained by solving the 1D scattering problem. It is assumed that  spin-up electrons ($\eta_z=1$) from the left lead inject into the first (left) FM island in Fig.~1(b) and then are scattered (note that for the opposite chirality $\eta_z=-1$ electrons, one should consider it injecting from the right lead), the scattering wavefunctions in each region are given by
\begin{equation}
\resizebox{.9\hsize}{!}{$\left\{
\begin{array}{c}
      \Psi_{\uppercase\expandafter{\romannumeral1}}(x<0)=
       \left(\begin{array}{c}
         \frac{1}{\sqrt{2}} \\
         0 \\
       \end{array}
     \right)
e^{ik_xx}+r_{\downarrow\uparrow}\left(
                     \begin{array}{c}
                      0 \\
                       \frac{1}{\sqrt{2}} \\
                     \end{array}
                   \right)
e^{-ik_{x}x} \\
   \Psi_{\uppercase\expandafter{\romannumeral2}}(0<x<L)=
    a_1\left(
       \begin{array}{c}
        M_{1x} \\
         u_1 \\
       \end{array}
     \right)
e^{i\kappa_{1}x}+b_1\left(
                     \begin{array}{c}
                             M_{1x} \\
                             v_1 \\
                     \end{array}
                   \right)
e^{-i\kappa_{1}x} \\
\Psi_{\uppercase\expandafter{\romannumeral3}}(L<x<L+L_0)=
    a_2\left(
       \begin{array}{c}
        1 \\
         0 \\
       \end{array}
     \right)
e^{ik_xx}+b_2\left(
                     \begin{array}{c}
                             0 \\
                             1 \\
                     \end{array}
                   \right)
e^{-ik_xx} \\
\Psi_{\uppercase\expandafter{\romannumeral4}}(L_0+L<x<2L+L_0)=
    a_3\left(
       \begin{array}{c}
        M_{2x} \\
         u_2 \\
       \end{array}
     \right)
e^{i\kappa_{2}x}\!+\!b_3\left(
                     \begin{array}{c}
                             M_{2x} \\
                             v_2 \\
                     \end{array}
                   \right)
e^{-i\kappa_{2}x}\\
   \Psi_{\uppercase\expandafter{\romannumeral5}}(x>2L+L_0)=
   t_{\uparrow\uparrow}\left(
                     \begin{array}{c}
                       \frac{1}{\sqrt{2}} \\
                       0 \\
                     \end{array}
                   \right)
e^{ik_xx}
  \end{array}
\right.$},
\end{equation}
where $\Psi_i$ ($i$=I-V) are the wavefunctions in the left lead, the left FM island, the normal $L_0$ region, the right FM island, and the right lead, respectively. $r_{\downarrow\uparrow}$ and $t_{\uparrow\uparrow}$ are the corresponding reflection and transmission amplitudes, $a_i$ and $b_i$ ($i$=1-3) are intermediate scattering coefficients. $u_{1,2}=E-\kappa_{1,2}$, $v_{1,2}=E+\kappa_{1,2}$, $k_x=E$, $\kappa_{1,2}=\sqrt{E^2-M_{1x,2x}^2}$ with $\hbar{v_F}=1$. The wavefunctions in each region are the superposition of eigenstates of local Hamiltonian. By matching wavefunctions at $4$ interfaces of the structure, we can get the following scattering coefficients $r_{\downarrow\uparrow}$  and $t_{\uparrow\uparrow}$ as
\begin{eqnarray}
r_{\downarrow\uparrow}=r_1+\frac{e^{i\varphi_0}t_1r_2t_1}{1-e^{i\varphi_0}r_1r_2}
\end{eqnarray}
and
\begin{eqnarray}
t_{\uparrow\uparrow}=t_1t_2e^{-2ik_{x}L}/(1-r_1r_2e^{i\varphi_0})
\end{eqnarray}
with $r_i=m_i(1-e^{2i\kappa_{i}L})/[(E+\kappa_{i})-e^{2i\kappa_{i}L}(E-\kappa_{i})]$,
$t_i=2\kappa_{i}e^{i\kappa_{i}L}/[(E+\kappa_{i})-e^{2i\kappa_{i}L}(E-\kappa_{i})]$ ($i=1,2$), $\varphi_0=2k_xL_0$.
Due to the definite chirality, the scattering coefficients $r_{\sigma\bar{\sigma}}$ and $t_{\sigma\sigma}$  are prohibitted ($\bar{\sigma}=-\sigma$), so the current formula can be rewritten as
\begin{eqnarray}
I_{L\sigma}=\frac{ie}{2\pi{T}}\oint_Td\tau\left(\frac{\partial r_{\sigma\bar{\sigma}}}{\partial \tau}r_{\sigma\bar{\sigma}}^*+\frac{\partial {t'}_{\sigma\sigma}}{\partial \tau}{t'}_{\sigma\sigma}^*\right),
\end{eqnarray}
where  the scattering coefficient $t'_{\sigma\sigma}$ is transmission of electrons from the right lead and tends to vanishing in a pump cycle when $E<E_{ef}$, whereas $|r_{\sigma\bar{\sigma}}|^2$ keep as a unit of $1$.  In Fig.~4(a), $I_{L\downarrow}$ is plotted as a function of the Fermi energy $E$, and the current-energy relationship is quite similar to those in Fig.~2 and Fig.~3, i.e. $I_{L\downarrow}$ and $I_{R\uparrow}$ are quantized in the energy gap and only one spin channel contributes to the pump current. Actually, other current-parameter relationships are fully the same (not shown). When $E$ is outside the energy gap, $E>E_{ef}$, the results in Fig.~4 are nonquantized and  smaller than $e/T$ different from those in Fig.~3(a). It is believed that such a distinction stems from  numerical calculations for Fig.~2 and 3. Since the numerics are based on a finite-size device and the finite pumping sites (sources) contributing to the pumping effect may lead to a much larger results of pumped currents due to the multiple quantum interferences.

\begin{figure}[ptb]
\centering
\includegraphics[bb=0 0 446 376, width=3.0in]{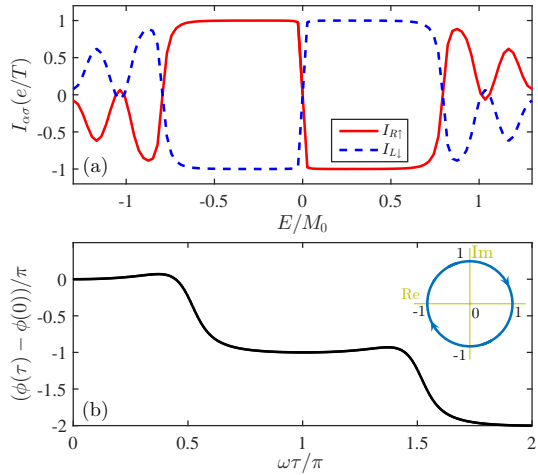}
\caption{(Color online) (a) Spin-dependent pumped current $I_{L\downarrow}$ and $I_{R\uparrow}$ as a function of the Fermi energy $E$ and (b) phase of the reflection coefficient $r_{\downarrow\uparrow}$ evolving with time $\omega\tau$.
The inset in (b) is the plot of phase trajectory of $r_{\downarrow\uparrow}$ in complex plane. Parameters are $\hbar{v_F=1}$, $E=0.004t$, $M_0=0.02t$, $L_0=0$, $L=200$.}
\end{figure}

To get some more insight into the quantized pump, we also plot the phase $\phi(\tau)$ of $r_{\downarrow\uparrow}$ as a function of $\omega\tau$ and its trajectory\cite{40} in Fig.~4(b). In our studied case, $\phi(\tau)$ decrements $2\pi$ in a cycle and the orbit of $r_{\downarrow\uparrow}$ is a unit circle on the complex plane. The trajectory is a closed orbit simply because the Hamiltonian is periodic in time and keeps invariant as long as $E<E_{ef}$. This indicates that the winding number of $r_{\downarrow\uparrow}$  is a unit of $1$ or $-1$ that corresponds to an integer number of charge pumped out through the system.

\section{Pure spin current}
From the above results, the two-parameter pump is able to generate a quantized charge current or fully spin-polarized charge current. However, it seems that the quantized pure spin current without any charge current, $I_s\neq0$, cannot be produced. There are some works\cite{38,39} verifying that a uniform magnetization precession on QSHI can lead to a topological spin pump. As is well known, the magnetization precession or Ferromagnetic resonance can pump out pure spin currents in usual metal or semiconductor devices, but where the spin currents are not quantized. Only in the topological materials can the magnetization gap the helical edge/surface states\cite{38,39}, the spin pump would be quantized. Nevertheless, we can also simulate such magnetization precession in our two-parameter pump device by considering two FM islands with perpendicular magnetization to each other: one is $M_{1x}=M_0\cos\omega\tau$ and the other is $M_{2y}=M_0\cos({\omega\tau+\varphi})$ in which the magnetization is along the $y$ axis, both of them keep altering with time and separated in real space.

We numerically calculate the double-boundary pump device [Fig.~1(a)] with two perpendicular FMs by using the lattice Hamiltonian of Eq.~(1). The counterparts of the single-boundary device are not shown here since it can be simply embodied in the former one. Parameters are taken the same as those in Fig.~2 but the right magnetization is set along the $y$ direction, $M_{2y}$. In Fig.~5(a), a nearly pure spin current is shown to flow through the device without a charge current ($I_{L\uparrow}=-I_{L\downarrow}$), and the current-phase relationship remains unchanged by comparing Fig.~5(a) with Fig.~2(b). This means that in each boundary of the pump device, the opposite spin is pumped out along the opposite direction. This situation is a little similar to the original pure spin current of helical edge states, however, the latter cannot automatically flow away from QSHI.

In Ref.~\cite{38} and \cite{39}, the spatially uniform magnetization precession was verified to generate a quantized pure spin current. While from our model, the two components of the magnetization precession, $M_{1x}$ and $M_{2y}$, separated in real space can also work to obtain the same results. There are some differences in physics origin behind these two methods. For a uniform magnetization precession, the pumping process is  that  a spin below/above the Fermi energy flows into the precession region and abosorbs/emits a photo energy $\hbar\omega$ to flip its spin, and then flows out system, so a spin current forms. Since the magnetization can open a gap of helical edge states, the pumped pure spin current remains the same when the Fermi energy resides in the energy gap, and we have $I(E)=I(-E)$. Actually, one can find that Eq.~(4) can be transformed exactly into the famous Rice-Mele model\cite{46} if the last two terms would be replaced by the precession term $M_{0}\sigma_x\cos\omega\tau+M_{0}\sigma_y\cos(\omega\tau+\varphi)$ in the spatially homogeneous system. So it is reasonable to get a quantized charge or spin pump, which depends on the single boundary or double boundaries of QSHI involved in the magnetization precession.

\begin{figure}[ptb]
\centering
\includegraphics[bb=0 0 446 560, width=3.0in]{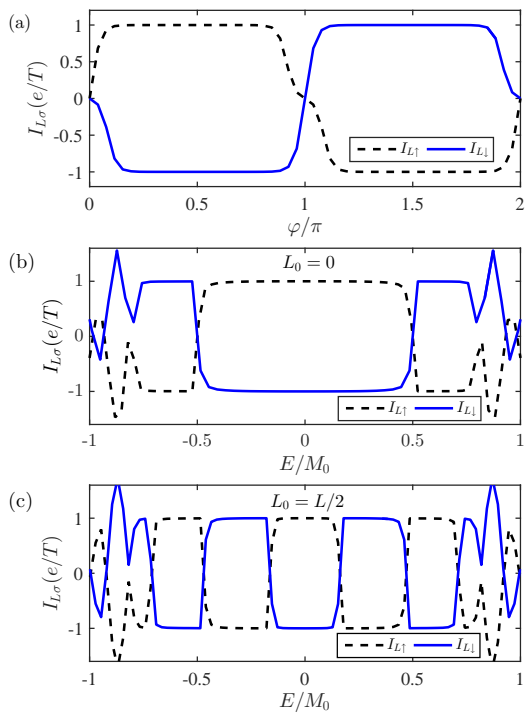}
\caption{(Color online) Plot of spin-dependent current $I_{L\sigma}$ in the double-boundary pump device as  functions of $\varphi$ in (a) and the Fermi energy $E$ in (b) and (c). Parameters are $E=0.001t$, $L=100$, $M_0=0.01t$, $L_0=0$ in (b), and $L_0=50$ in (c).}
\end{figure}

For our studied model, the pumping phase difference $\varphi$ plays a decisive role in controlling currents, $I_L=0$ at $\varphi=n\pi$ as shown in Fig.~5(a). Furthermore, the two-parameter charge pump  stems in essence from the quantum interference effect\cite{45}, so when the dynamic phase of traveling  particles in device alters with an increase of $L_0$ in Fig.~5(b) and 5(c), the spin current direction would be reversed periodically although it keeps quantized $I_{L\uparrow}=-I_{L\downarrow}=\pm{e/T}$. In addition, the particle-hole antisymmetry seems destroyed from the single $I_{L\uparrow}$ or $I_{L\downarrow}$, because the introduced pump parameters on the helical edge states, $M_{1x}$ and $M_{2y}$,  destroy the chiral symmetry of system, i.e., $I_{L\uparrow/\downarrow}(E)\neq-I_{L\uparrow/\downarrow}(-E)$ in Fig.~5(b) and ~5(c). Nevertheless, we have the relationship $I_{L\uparrow/\downarrow}(E)=I_{L\uparrow/\downarrow}(-E)$. Actually, the pumped pure spin current not the charge current has the particle-hole symmetry similar to the case of the uniform magnetization precession.

\section{Conclusion}
In summary, we have investigated  possible two-parameter quantized pump based on the helical edge states of a two-dimensional topological insulator. Taking two time-dependent magnetizations as pumping potentials with a phase difference between them, we  in both numerical and continuum models showed that a quantized charge or spin pump is available. The pumping quantization is due to the time-dependent magnetization that opens an energy gap of the original material to form a new topological interface state, and thus is protected by the topology. It is also found that the quantized current can be fully-spin polarized, unpolarized, or pure spin current. The current direction can be reversed abruptly by system parameters such as the Fermi energy, the pumping phase, and the local static potential. Our findings may pave a new way to generate quantized spin pump in a two-parameter pump device.

\begin{acknowledgments}
The work is supported by NSFC (Grant Nos. 11574045, 11774144). We are very grateful to Dr. L. Sheng for helpful discussions.
\end{acknowledgments}

\end{document}